\documentclass[12pt]{article}
\newcommand{\dd}{\mbox{\rm d}}

\newcommand{\oo}{\over}
\newcommand{\p}{\partial}

\newcommand{\be}{\begin{equation}}
\newcommand{\bear}{\begin{eqnarray}}
\newcommand{\ear}{\end{eqnarray}}
\newcommand{\ee}{\end{equation}}

\newcommand{\bi}{\bibitem}
\newcommand{\ci}{\cite}

\newcommand{\vs}{\vspace}

\begin{document}
\thispagestyle{empty}

\vs{2cm}

\begin{center} {\LARGE \bf An Alternative to Matter Localization in
the ``Brane World": An Early 

Proposal and its Later Improvements}

\vspace{1cm}
       
Matej Pav\v si\v c

Jo\v zef Stefan Institute, Jamova 39, SI-1000 Ljubljana, Slovenia;

e-mail: matej.pavsic@ijs.si

\vs{3mm}

January 2001

\vspace{2cm}

\end{center}

Here we place the Latex typeset of the paper\footnote{This rather old
paper is not available elsewhere on the Internet. Therefore, facing the
explosion of the activity concerning the brane world, I would like
to make the paper easily accessible. The body of the paper is identical
to the published version; a footnote is added, and a note at the end.}

\vs{1mm}

M. Pav\v si\v c, Phys. Lett. A {\bf 116}, 1--5 (1986)

\vs{1mm}

In the paper we presented the picture that our spacetime is a 3-brane
moving in a higher dimensional space. The dynamical equations were
derived from the action which is just that for the usual Dirac--Nambu--Goto
$p$-brane. We also considered the case where not only one, but many
branes of various dimensionalities are present, and showed that their 
intersections with the 3-brane manifest as
matter in 4-dimensional spacetime. We considered a particular case, where the
intersections behaved as point particles, and found out that they
follow the geodesics on the 3-brane worldsheet (identified with our spacetime).
In a series of subsequent papers the original idea has been further improved and
developped. This is discussed in a note at the end, where it is also
pointed out that such a model resolves the problem of 
massive matter confinement
on the brane, recently discussed by Rubakov et al. and M\" uck et al.\,.

\newpage
\thispagestyle{empty}

{\it Physics Letters A} {\bf 116}, 1--5 (1986)
 
\vspace{3cm}

\begin{center} {\LARGE \bf Einstein's Gravity from a First Order
Lagrangian in an Embedding Space}

\vspace{1cm}
       
Matej Pav\v si\v c

Jo\v zef Stefan Institute, University of Ljubljana, Yugoslavia

\vs{3mm}

19 May 1986

\vspace{2cm}

\end{center}

\vspace{.8cm}

We formulate a first order action principle in a higher dimensional space
$M_N$ in which we embed spacetime. The action $I$ is essentially an
``area" of a four-dimensional spacetime $V_4$ weighted with a matter
density $\omega$ in $M_N$. For a suitably chosen $\omega$ we obtain
on $V_4$ a set of worldlines. It is shown that these worldlines are
geodesics of $V_4$, provided that $V_4$ is a solution to our variational
procedure. Then it follows that our spacetime satisfies the Einstein equations
for dust -- apart from an additional term with zero covariant divergence.
(This extra term was shown in a previous paper to be exactly zero at least
in the case of the cosmolical dust model.) Thus we establish a remarkable 
connection of the extrinsic spacetime theory with the intrinsic
general relativity. This step appears to be important for quantum gravity.

\newpage

The idea of formulating a theory of gravity by using an embedding space 
has been issued by some authors such as Fronsdal \cite{1} and
others \ci{2}. So far it has not received much attention. A possible reason
is that the Einstein general relativity based on the intrinsic geometry
of spacetime is so successful. However, if we try to quantize the theory
we face serious difficulties \ci{3} which are only partially circumvented
by the brilliant works of many authors \ci{4}--\ci{6}. Therefore it seems
reasonable to search for extensions and/or alternatives to the Einstein
gravity. A promising approach appears to be a reformulation of general
relativity in the extrinsic instead of the intrinsic terms only.

In the previous papers \ci{7,8} we started to develop an idea which is
briefly the following. We assume that the arena, in which the physical
events (parametrized by the coordinates $\eta^a$, $a - 1,2,...,N$) are
situated, is a certain higher dimensional pseudoeuclidean space $M_N$ --
{\it with a given dimension} $N$ (say 10). With these events we associate
a field $\omega (\eta)$ which represents what I call {\it the matter
density} in $M_N$. This higher dimensional world is ``frozen" and
motionless: everything is written in $M_N$ once and for all.

Then we assume that somewhere in $M_N$ the matter density is ``wired"
in such a way that it allows for the existence of 
{\it an observer}\footnote{Our observer by definition possesses the property
called ``{\it consciousness}" which in my terminology means that he not
only ``registers" the outside events, but he registers also the fact that
he has registered, etc. ad ``infinitum". So we have a succession of
registrations, and memory of what has benn registered, and so on.
Each registration together with the registration of registration, etc., we
shall name {\it observation}. The succession of observations could be named
``stream of conscioussness".}. By assumption, an observer registers
(in classical approximation), at each step, only the events on a
three-dimensional surface $\Sigma$; at the next step on some other
three-surface $\Sigma'$, etc.\footnote{{\bf Note added in January, 2001:}
This has to be taken with care. At a given moment of his proper time the
observer, of course, does not register all the events on a surface $\Sigma$.
Because of the finite speed of light, he is only able to accumulate data on
$\Sigma$ with the passage of his proper time and thus gradually expand
the region of $\Sigma$ known to him. Hence, he can talk about ``motion"
of the (simultaneity) surface $\Sigma$, but the latter is becoming known
to him a posteriori. See also the note at the end.}
So the observer by the very act of successive
observations introduces {\it the motion} of his simultaneity three-surface
$\Sigma$ through the higher space $M_N$. The set of all three-surfaces
$\Sigma$ belonging to the series of observations of our observer forms
a four-dimensional continuum $V_4$, called {\it spacetime}\footnote{
Since within the same wired structure ${\cal B}$ in $M_N$ supporting
the existence of an observer ${\cal O}$ we could start from various
three-surfaces $\Sigma$ which when moving would describe various
spacetimes $V_4$, we must provide a mechanism which would pay attention to
one particular $V_4$ only. This is achieved through {\it the process of
successive observations -- picturesquely called ``stream of consciousness" --}
as defined in footnote 1. Various streams of consciousness are possible
with the same structure ${\cal B}$ and they belong to various observers
${\cal O}$ on different spacetimes ``going" through ${\cal B}$.
The registrations alone as performed for instance by a measuring
apparatus and stored in a magnetic tape) are not sufficient to determine
a particular $V_4$, that is a path of a three-surface $\Sigma$
through $M_N$.},
parametrized by the coordinates $x^{\mu}$ ($\mu = 0,1,2,3$).

A spacetime $V_4$ is represented by the equation $\eta^a = \eta^a (x)$,
whilst the induced metric tensor on $V_4$ is $g_{\mu \nu} = \p_{\mu}
\eta^a \, \p_{\nu} \eta_a$. In a classical theory we assume that dynamically
allowed are only such $V_4$'s which are solutions of a certain variational
principle in $M_N$. In refs. \ci{7,8} we have taken the second order 
Einstein--Hilbert action $I_H$ expressed in terms of the extrinsic
variables $\eta^a$, $\p_{\mu} \eta^a$, $\p_{\mu} \p_{\nu} \eta^a$ plus
a corresponding matter action $I_m$. Varying $I_2 = I_H + I_m$ with
respect to $\eta^a$ we have obtained the covariant divergence of
Einstein's equations multiplied by $\p_{\nu} \eta_a$. These equations imply the
validity of Einstein's equations with an additional term $C^{\mu \nu} (x)$,
such that $(C^{\mu \nu} \p_{\nu} \eta_a)_{;\mu} = 0$ and 
${C^{\mu \nu}}_{;\nu } = 0$.
Then we argued that in order to fix a solution (i.e., $V_4$) we need some
additional equation (besides a choice of gauge). We have chosen 
the equation that results by varying the first order action\footnote{
{\bf Note added in January, 2001 :} Such an action can be straightforwardly
derived from the Dirac--Nambu--Goto action in a conformally flat embedding
space \ci{a0}.}
\be
      I = \int \omega \sqrt{-g} \, \dd^4 x = I[\eta^a (x)] ,
\ee
where $\omega$ is a function of position $\eta^a$ in the embedding space
and $g$ the determinant of the intrinsic metric. Finally we have concluded
that our starting action for the ``field" $\eta^a (x)$ is $I$ of
eq.\, (1), whilst the equation resulting from the second order action $I_2$
served served for the purposes of calculating the four-velocity $u^{\mu}$.
In the present work I am going to demonstrate that we can avoid the use
of $I_2$ and that the action (1) -- for a suitable $\omega$ -- already
contains the Einstein equations for dust, apart from a function $C^{\mu \nu}
(x)$ satisfying ${C^{\mu \nu}}_{; \nu} = 0$.

If we vary $\eta^a (x)$ in the action (1) we obtain \ci{7,8} (see also
\ci{9})
\be
        {1 \oo \sqrt{-g}} \p_{\mu} (\sqrt{-g} \, \omega \p^{\mu} \eta_a ) =
        \p_a \omega \quad ; \qquad \p_a \equiv {\p \oo {\p \eta^a}} .
\ee
For a given $\omega (\eta)$, initial and boundary conditions and a chosen
parametrization $x^{\mu}$, we obtain a unique solution $\eta^a (x)$
describing a spacetime $V_4$.

The matter density on a particular spacetime $V_4$ (represented by
$\eta^a = \eta^a (x)$) is
\be
   \omega (\eta (x)) \equiv \rho (x)
\ee
and depends on the choice of $\omega (\eta)$ and of $\eta^a (x)$.

In the following we shall assume that the matter density $\omega (\eta)$
consists of massive $m$-dimensional sheets ${\hat V}_m^{(i)}$, parametrized
by ${\hat x}^A$ ($A = 1,2,...,m$) and represented by the equation 
$\eta^a = {\hat \eta}_i^a ({\hat x})$; to this we also add a constant
$\omega_0$:
\be
    \omega (\eta) = \omega_0 + \int \sum_i m_i \, \delta^N (\eta - 
    {\hat \eta}_i) \dd^m {\hat x} .
\ee

From the point of view intrinsic to a particular $V_4$ (already given as a
solution of eq.\, (2)), the matter density $\omega (\eta)$ of eq.\, (4)
becomes (up to $\omega_0$) that of point particles:
\be
   \omega (\eta (x)) \equiv \rho (x) = \omega_0 + \int \sum_i m_i
   {1\oo \sqrt{-g}} \delta^4 (x - z_i) \dd s_i ,
\ee
provided that we choose $m = N - 4 + 1$ (for $N = 10$ we have $m =7$).
That is, the intersection $V_4 \cap \, \bigcup_i {\hat V}_m^{(i)}$ gives (5).
In other words, when crossing $V_N$ with a spacetime $V_4$, then, using (4)
we obtain on $V_4$ the set of worldlines $x^{\mu} = z_i^{\mu} (\lambda)$;
on a three-surface $\Sigma$ this is the matter distribution of 
point particles, forming a so called {\it dust}.

Inserting (5) into (1) we have, apart from the term $I_0 = \int \omega_0 \,
\sqrt{-g} \dd^4 x$, the action of point particles in a given gravitational
field $g_{\mu \nu}$:
\be
    I = I_0 + \int \sum_i m_i \, \dd s_i = I_0 + \int \sum_i m_i \,
    {(g_{\mu \nu} {\dot z}_i^{\mu} {\dot z}_i^{\nu})}^{1/2} \, \dd \lambda_i ,
\ee
where $\dd s_i$ is a proper time and $\lambda_i$ and arbitrary parameter along
the $i$th worldline.

It is really remarkable that the gravitational field $g_{\mu \nu} (x)$ in
{\it the intrinsic} action (6) is actually determined by the variation of the
original action in $V_N$ with respect to $\eta^a (x)$ (eqs.\,(1) and (4)).
Usually -- when starting from the intrinsic point of view -- the action
(5) describes (dust) particles in a {\it fixed} field $g_{\mu \nu} (x)$;
varying $z_i^{\mu}$ we obtain the geodesic equation. To obtain the equations
of motion for $g_{\mu \nu}$ we need an additional term, namely $(16 \pi G)^{-1}
\int \sqrt{-g} \dd^4 x \, R$; this leads to the Einstein equations
$G^{\mu \nu} = - 8 \pi G \rho u^{\mu} u^{\nu}$ which -- because of the
contracted Bianchi identity $(\rho u^{\mu} u^{\nu})_{; \nu} = 0$ and the
relation $(\rho u^{\nu})_{;\nu} = 0$ for dust -- imply the geodesic
equation ${u^{\mu}}_{; \nu} u^{\nu} = 0$.

Now, from {\it the extrinsic point of view} (extrinsic relative to $V_4$) no
additional terms are needed: the action (1) together with a propersly chosen
$\omega (\eta)$, in our case that of eq.\,(4), suffices both for the
determination of a spacetime $V_4$ and a ``motion"\footnote{In our
philosophy an observer's stream of consciousness (see footnote 2) is
actually moving (this implies that a spacelike three-surface $\Sigma$
is moving), whilst matter is motionless. This $\Sigma$-motion leads 
to the illusion that point particles are moving in three-space. In four-space
$V_4$ we have worldlines and point particles are three sections of
worldlines; in still higher space $M_N$ matter is represented by
$m = N-3$ sheets ${\hat V}_m$ and worldlines are intersections with
a chosen spacetime $V_4$.} of particles in it.

However, it is not obvious that the equations for $\eta^a (x)$ (eq.\,(2))
contain the geodesic equation for a worldline. This I shall now explicitly
demonstrate.

Let us integrate (2) over $\sqrt{-g} \, \dd^4 x$ on a chosen $V_4$ and let
the domain of integration $\Omega$ embrace one worldline only. The we have
\be
    \int_{\Omega} \p_{\mu} (\sqrt{-g} \, \omega \, \p^{\mu} \eta_a) \, \dd^4 x
    = \oint_B \sqrt{-g}\, \omega \, \p^{\mu} \eta_a \, \dd \Sigma_{\mu}
    = \int_{\Omega} \p_a \omega \sqrt{-g} \, \dd^4 x ,
\ee    
where $\dd \Sigma_{\mu}$ is an element of a three-surface $B$. The 
integration over
the constant part of $\omega = \omega_0 + \omega_1$ can be made 
arbitrarily small, 
since for a point worldline we can shrink $\Omega$ to zero.

The boundary $B$ in (7) consists of two spacelike three three-surfaces
$\Sigma_1$ and $\Sigma_2$ and a timelike three-surface; integration over the 
last one gives zero, since $\omega_1$ is zero around a worldline. Then,
writing $\dd \Sigma_{\mu} = n_{\mu} \dd \Sigma$, $U_a \equiv \p^{\mu}
\eta_a n_{\mu}$ and $\dd^4 x = \dd \Sigma \, \dd s$ we have from (7)
\be
    \int_{\Sigma_1}^{\Sigma_2} \sqrt{-g} \omega_1 U_a \, \dd \Sigma =
    m_i U_a \vert_{\Sigma_2} - m_i U_a \vert_{\Sigma_1} =
    \int \p_a \omega \sqrt{-g} \, \dd \Sigma \, \dd s_i .
\ee 

If $\Sigma_1$ and $\Sigma_2$ are infinitesimally close, then (8) becomes
\be
m_i \, {{\dd U_a}\oo {\dd s_i}} = {\dd \oo {\dd s_i}} \int \p_a \omega_1 
\sqrt{-g} \, \dd^4 x = \int \p_a \omega_1 \sqrt{-g} \, \dd \Sigma ,
\ee
where $m_i$ is the $i$th particle mass, $n_{\mu}$ the normal to $\Sigma$,
and $\dd s_i$ the element of proper time along the $i$th worldline.

Suppose that we have solved eq.\,(2) and found $\eta^a (x)$ for the case of 
$\omega (\eta)$ such as given in (4). The four-surface $\eta^a (x)$\footnote{
Here $\eta^a (x)$ is a representative of the class $\lbrace \eta^a (x)
\rbrace$ of $\eta^a$'s which all describe the same spacetime $V_4$ using
different parametrizations.} near the $i$th matter sheet ${\hat V}_7^{(i)}$
is affected by the presence of all the sheets ${\hat V}_7$ including one
of the $i$th particle. A matter sheet ${\hat V}_7$ is supposed to influence
$V_4$ in two ways
\begin{description}
\item{i)} it ``twists" $V_4$ {\it without changing} (up to a reparametrization)
{\it the intrinsic metric} $g_{\mu \nu} (x)$ in the vicinity of a 
worldline $C_i$;

\item{ii)} it additionally reshapes $V_4$ so that $g_{\mu \nu}$ (and curvature)
close enough to $C_i$ is also changed.

\end{description}

We repeat that a worldline $C$ is the intersection of a spacetime $V_4$ 
and a matter sheet ${\hat V}_7$. Different spacetimes give different
worldlines $C$ for the same matter sheet ${\hat V}_7$. These different
$V_4$ may all have the same intrinsic metric $g_{\mu \nu} (x)$ (and the
curvature) everywhere except sufficiently close to $C$. Let us therefore
write $\eta^a (x)$ as the sum of ${\tilde \eta}^a (x)$ due to the effect (i)
and $\eta^{*a} (x)$ due to (ii):
\bear
        \eta^a (x) &=& {\tilde \eta}^a (x) + \eta^{*a} (x) \nonumber \\
        \p_{\mu} \eta^a &=& \p_{\mu} {\tilde \eta}^a + \p_{\mu} \eta^{*a} .
\ear
Then we have
\be
      U^a = n^{\mu} \p_{\mu} \eta^a = n^{\mu} (\p_{\mu} {\tilde \eta}^a +
      \p_{\mu} \eta^{*a} ) \equiv {\tilde U}^a + U^{* a} .
\ee

The derivative $\p_{\mu}$ can be split into the normal (relative to 
a spacelike three-surface $\Sigma$) derivative ${\hat \p} \equiv n^{\mu} 
\p_{\mu}$ and the tangential derivative ${\bar \p}_{\mu}$ \ci{10}. Then
\be
     U^{*a} \equiv n^{\mu} \p_{\mu} \eta^{*a} = n^{\mu} (n_{\mu}
     {\hat \p} \eta^{*a} + {\bar \p}_{\mu} \eta^{*a}) = {\hat \p} \eta^{*a} .
\ee
Since we can choose $\Sigma$ so that on a worldline $C$ the normal $n^{\mu}$
coincides with the tangent $u^{\mu}$ of $C$, it holds that ${\hat \p}
\eta^{*a}$ points into the direction of the worldline. Since worldline's 
density is the same everywhere on $C$, it follows from the symmetry 
consideration that $\eta^{*a} (x)$ is constant along a worldline and
therefore ${\hat \p} \eta^{*a} = 0$. Hence
\be
    U^a = {\hat U}^a \equiv n^{\mu} \p_{\mu} {\tilde \eta}^a .
\ee

Let us multiply (9) by $\p^{\mu} {\tilde \eta}_a$. Since in
${\tilde \eta}^a (x)$ we exclude the contribution of the $i$th particle to
$g_{\mu \nu}$ (though we still retain the influence of the $i$th 
${\hat V}_7^{(i)}$ to the extrinsic reshaping of $V_4$) it holds that
$\p^{\mu} {\tilde \eta}^a$ remains practically constant over the small 
four-region $\Omega$ surrounding our worldline. Therefore we have from (9)
\bear
   m_i \, {{\dd U_a}\oo {\dd s_i}} \p^{\mu} {\tilde \eta}^a &=&
   \p^{\mu} {\tilde \eta}^a \int \p_a \omega_1 \sqrt{-g} \dd \Sigma \nonumber \\
   &\approx& \int \p^{\mu} {\tilde \eta}^a \p_a \omega_1 \sqrt{-g} \dd \Sigma
   \nonumber \\
   &=& \int \p^{\mu} \omega_1 \sqrt{-g} \dd \Sigma
\ear
Since \ci{7}--\ci{9} $U_a ={\tilde U}_a = \p_{\nu} {\tilde \eta}_a e^{\nu}$,
$\p_{\alpha} \p_{\beta} {\tilde \eta}_a \p^{\rho} {\tilde \eta}^a =
{\tilde \Gamma}_{\alpha \beta}^{\rho}$ and $\p^{\mu} {\tilde \eta}^a 
\p_{\nu} {\tilde \eta}_a = {\delta^{\mu}}_{\nu}$, where 
${\tilde \Gamma}_{\alpha \beta}^{\rho}$ is the affinity of $V_4$ (described
by ${\tilde \eta}^a (x)$) and $u^{\nu}$ is $n^{\nu}$ on the worldline, it
is
\be
    {{\dd U_a}\oo {\dd s}} \, \p^{\mu} {\tilde \eta}^a = {{\dd u}\oo {\dd s}}
    + {\tilde \Gamma}_{\alpha \beta}^{\mu} u^{\alpha} u^{\beta} .
\ee

For the purpose of calculating the right-hand side of eq.\,(14) it is 
convenient to replace the $\delta$-function in $\omega_1$ by its finite,
analogue, e.g., a normalized Gaussian function of width $\sigma$; the
$\delta$-function is then a limit for $\sigma \rightarrow 0$.

Let us separate
\be
    \p^{\mu} \omega_1 = n^{\mu} {\hat \p} \omega_1 + {\bar \p}^{\mu} \omega_1 .
\ee
Since a worldline is the same at all values of proper time $s$, it
follows that the derivative of $\omega_1$ along a worldline is zero:
${\hat \p} \omega_1 =0$. The second term in (16) can be written in a
special coordinate system as $\p^r \omega_1$ with $r = 1,2,3$. The right-hand
side of (14) is then $\int \p^r \omega \sqrt{-g} \, \dd \Sigma$ and it obviously
vanishes if the domain of integration encloses a particle.

To be specific we now assume that
\be
   \omega_1 (\eta (x)) \equiv \rho (x) - \omega_0 =
   \int \sum_i m_i A(\sigma) {\rm exp} \left [ - {{(x - z_i)^2}\oo {2 \sigma^2}}
   \right ] \, \dd s_i ,
\ee
where $A(\sigma)$ is a suitable normalization constant. In the limit
$\sigma \rightarrow 0$ this becomes the sum of the $\delta$-functions as
given in (5). Using (17) the intergal (14) becomes
\be
    {\dd \oo {\dd s_i}} \int_{\Omega_i} \p^{\mu} \omega_1 \sqrt{-g} \, 
    \dd^4 x = {\dd \oo {\dd s_i}} \int_{\Omega_i} m_i A(\sigma)
    (x^{\mu} - z_i^{\mu}) \, {\rm exp} \left [ - {{(x - z_i)^2}\oo {2 \sigma^2}}
   \right ] \sqrt{-g} \, \dd^4 x = 0 .
\ee
where the integration runs over a four-region $\Omega_i$ surrounding a point
$z_i^{\mu}$ on a worldline $C_i$.

Therefore, eq.\,(14), which we obtained from the equation for $\eta^a (x)$
resulting from our first order variational principle (1), becomes the
geodesic equation:
\be
{{\dd u}\oo {\dd s}}
    + {\Gamma}_{\alpha \beta}^{\mu} u^{\alpha} u^{\beta} = {u^{\mu}}_{;\nu}
    u^{\nu} = 0 ,
\ee
where we omit ``$\sim$" which distinguishes $V_4$ from ${\tilde V}_4$,
since both spacetimes coincide everywhere except sufficiently near 
the worldline.

The result (19) is fascinating, because the same result also follows from
Einstein's equations for dust. Is the opposite also true, namely that (2)
and (4) which lead to (19) imply Einstein's equations for an induced metric?

Since in our case we also have $(\rho u^{\nu})_{;\nu} = 0$, it follows
from (19) that
\be
     (\rho u^{\mu} u^{\nu})_{;\nu} = 0 .
\ee

On the other hand our metric identically satisfies the contracted Bianchi
idintities
\be
      {G^{\mu \nu}}_{;\nu} = 0 ,
\ee
where $G^{\mu \nu} \equiv R^{\mu \nu} - {1\oo 2} g^{\mu \nu} R$ is the
Einstein tensor. Eqs.\,(20) and (21) imply the existence of a function
$C^{\mu \nu} (x)$ satisfying
\be
     G^{\mu \nu} = - 8 \pi G (\rho u^{\mu} u^{\nu} + C^{\mu \nu}) =
     - 8 \pi G T^{\mu \nu} ,
\ee
where $G$ is a suitable constant, e.g., the gravitational constant. Remember
that in our case $G^{\mu \nu}$, $\rho$ and $u^{\mu}$ are already given by
a solution $\eta^a (x)$ of eqs.\,(2) and (4). Therefore eq.\,(22) is a
defining equation for $C^{\mu \nu} (x)$. It measures a deviation of the
proposed dust model (eqs.\,(2) and (4)) from Einstein's equations for
dust: the stress energy tensor $T^{\mu \nu}$ has the term for dust and
the additional term $C^{\mu \nu}$. Though the induced metric and the
four-velocity obtained from eqs\,(2) and (4) in general is not expected to
satisfy the Einstein equations for dust, it is nevertheless significant that
the deviative term $C^{\mu \nu}$ has zero covariant divergence and -- as we
have seen -- the dust particles follow geodesics. In a previous work I have
shown that there exists at least one case (namely the cosmological dust
model) in which $C^{\mu \nu}$ is exactly zero.

It is really fascinating and important also for quantization that such result
follows from the simple first order action (1). However, for the sake 
of completeness, 
it would be necessary to solve explicitly eq.\,(2) for $\eta^a (x)$, find
the intrinsic metric in the vicinity of a point particle and compare
it with the Schwarzschild solution of the Eisntein equations. Analogous
has been already done for a {\it continuous} $\omega$ \cite{7,8}, and found
that the situation is exactly as predicted by the cosmological dust
solution of the Einstein equations.

Here we have considered, as a particular case, the intersection 
-- in $M_{10}$ --
of $V_4$ with ${\hat V}_7$ which, in general, gives a worldline on $V_4$.
If instead of ${\hat V}_7$ we take ${\hat V}_8$, then we obtain a worldsheet
on $V_4$, that is {\it a string} in three-space. It would be interesting
to investigate also into this direction, especially in view of recent
successes in string theory.

\vs{2mm}

\newpage

{\bf \Large Note added in January, 2001}

\vs{3mm}

In a series of papers \ci{a1}--\ci{a6} the model has been further ellaborated.
In ref.\,\ci{a2} a more direct proof that the worldline resulting from
the intersection of two branes is a geodesic on the brane worldsheet
has been found. In \ci{a1,a2} it was pointed out, following the known
results of refs.\,\ci{b1}, that the effective (4-dimensional)
gravity on the brane is that of Einstein plus the higher order corrections.

In ref.\,\ci{a1}--\ci{a3} a conceptual shift has been gradually done,
culminating in ref.\,\ci{a4}, so that -- in order to make the idea more
along the lines with the current practice in physics --  I did no longer 
worry about the observer, observation, succession of observations, etc.\,.
Instead, I considered simply the extended objects, $p$-branes,
living in spacetime which -- according to the usual theory -- has
necessarily more than 4 dimensions. Then I proposed that, instead of
trying to compactify those extra dimensions, it seem more naturally to
assume that a 3-brane worldsheet $V_4$ already represents our
spacetime. No compactification of the embedding space (also called
target space) is necessary.
In a recent preprint \ci{a6} I considered the case where many branes
were present. All the branes were considered as dynamical, their equations
of motion being derived from the corresponding action. The branes may
intersect or self-intersect.
It was shown that {\it the brane intersections and self-intersections
behave as matter in the brane world}. This resolves the problem of matter
confinement on the brane as pointed out recently by M\" uck, et al. \ci{a7},
who found that massive particles cannot stably move along the brane:
the brane is repulsive, and matter will be expelled from the brane into the
extra dimensions. A similar behavior was observed by Dubovsky and Rubakov
\ci{a8} in a different context.

In refs.\,\ci{a1}--\ci{a3} various technical and conceptual aspects\footnote{In those
papers the issue raised in footnote 3 is ellaborated further.}
of the idea that our world is a 3-brane living in a higher dimensional
space were investigated, including the possibility of resolving the
notorious ``problem of time" in quantum gravity. A similar resolution,
though not within the brane world context, has been proposed in 
ref.\,\ci{a9}.

\end{document}